\documentclass[aps,prl,reprint,amsmath,amssymb,superscriptaddress,showkeys]{revtex4-1}
\pdfoutput=1

\usepackage[latin1]{inputenc}
\usepackage{graphicx}
\usepackage{float}
\usepackage{amsmath}
\usepackage{amsfonts}
\usepackage{amsmath}
\usepackage{amssymb}
\usepackage[dvipsnames]{xcolor}
\usepackage{amsthm}
\usepackage{mathtools}
\usepackage{bm}
\usepackage{dsfont}
\usepackage{stmaryrd}
\usepackage{xspace}
\usepackage{epstopdf}
\usepackage{makecell}
\usepackage{siunitx}
\usepackage[shortlabels]{enumitem}
\usepackage{hyperref}

\newcommand\myshade{85}
\colorlet{mylinkcolor}{BrickRed}
\colorlet{mycitecolor}{NavyBlue}
\colorlet{myurlcolor}{Aquamarine}

\hypersetup{
  linkcolor  = mylinkcolor!\myshade!black,
  citecolor  = mycitecolor!\myshade!black,
  urlcolor   = myurlcolor!\myshade!black,
  colorlinks = true,
}

\graphicspath{ {img/} }

\definecolor{BrewerRed}{RGB}{228,26,28}
\definecolor{BrewerBlue}{RGB}{55,126,184}

\DeclareMathOperator{\E}{\mathds{E}}

\let\oldparagraph\paragraph
\renewcommand\paragraph[1]{\vskip0.5cm\oldparagraph{#1}}

\newcommand\cX{\ensuremath{\mathcal{X}}\xspace}
\newcommand\cY{\ensuremath{\mathcal{Y}}\xspace}

\newcommand\subref[2]{\hyperref[#1]{\ref*{#1}#2}}

\begin{document}

\title{Decomposing spectral and phasic differences in non-linear features between datasets}

\author{Pedro A.M. Mediano}
\thanks{P.M. and F.R. contributed equally to this work.\\E-mail: pam83@cam.ac.uk, f.rosas@imperial.ac.uk}
\affiliation{Department of Psychology, University of Cambridge, Cambridge CB2 3EB}

\author{Fernando E. Rosas} 
\thanks{P.M. and F.R. contributed equally to this work.\\E-mail: pam83@cam.ac.uk, f.rosas@imperial.ac.uk}
\affiliation{Center for Psychedelic Research, Department of Medicine, Imperial College London, London SW7 2DD}
\affiliation{Data Science Institute, Imperial College London, London SW7 2AZ}
\affiliation{Center for Complexity Science, Imperial College London, London SW7 2AZ}

\author{Adam B. Barrett}
\thanks{A.B.B. and D.B. are joint senior authors.}
\affiliation{Sackler Center for Consciousness Science, Department of Informatics, University of Sussex, Brighton BN1 9RH}

\author{Daniel Bor}
\thanks{A.B.B. and D.B. are joint senior authors.}
\affiliation{Department of Psychology, University of Cambridge, Cambridge CB2 3EB}

% \date{\today}

\begin{abstract}

When employing non-linear methods to characterise complex systems, it is
important to determine to what extent they are capturing genuine non-linear
phenomena that could not be assessed by simpler spectral methods. Specifically,
we are concerned with the problem of quantifying spectral and phasic effects on
an observed difference in a non-linear feature between two systems (or two
states of the same system). Here we derive, from a sequence of null models, a
decomposition of the difference in an observable into spectral, phasic, and
spectrum-phase interaction components. Our approach makes no assumptions about
the structure of the data and adds nuance to a wide range of time series
analyses.

\end{abstract}

\maketitle

Non-linear methods are useful for characterising differences between various
states of a complex system, and have found applications in a wide range of
scientific domains. For example, Lempel-Ziv (LZ) complexity~\cite{Lempel1976}
and multiscale entropy~\cite{Costa2002} have been successful in discriminating
between conscious and unconscious brain activity~\cite{Casali2013}, and have
yielded insights into physiological pathologies~\cite{costa2003multiscale} and
price dynamics~\cite{Martina2011}. However, more refined conclusions could be
obtained if there were a principled way to assess how much of the differences
in such measures are due to genuine non-linear effects, and how much is
explainable by changes in the power spectrum.

A popular approach to study the effect of spectral and phasic contributions on
an observable is via \emph{surrogate data methods}~\cite{Theiler1992}, which
examine whether its value is representative of a null distribution obtained
from surrogate data. Such surrogate methods are regarded as a basic constituent
of the data analyst's toolkit~\cite{Lancaster2018}, and have been extended to a
range of scenarios including multivariate time series~\cite{Prichard1994},
non-stationary data~\cite{Lucio2012}, and many others. However, surrogate
methods are typically designed to be applied on \emph{a single dataset}, and it
is not straightforward to use them to disentangle spectral and phasic
contributions on differences in an observable \emph{between two datasets} ---
e.g. how much of the difference in LZ complexity between two neurological
conditions simply reflects the known spectral changes between
them~\cite{Schiff2014}. The crux of why this is challenging, and why naive
applications of typical surrogate methods fail, is that \emph{the difference
between two null models is not necessarily a good null model of the difference}
(see Supp. Mat. for a detailed example).

To deal with this issue, here we present a novel decomposition of the
difference in an observable between two time series datasets into spectral,
phasic, and spectrum-phase interaction components. The decomposition makes no
assumptions about the structure of the data, and is widely applicable to a
broad range of scenarios of interest. We illustrate our method by analysing LZ
complexity on neuroimaging data, where our decomposition identifies phasic and
spectrum-phase interaction components that take the opposite sign to the
predominantly spectral overall effect, and which would not have been detectable
by previously existing methods.

\paragraph{The decomposition.}

Let us consider a scientist who is interested in an observed difference in some
quantity $f$ between data recorded in two different conditions, denoted by
$\mathcal{X}$ and $\mathcal{Y}$. The data consist of time series recordings,
and a set of time series segments are obtained from each condition. Each
segment could correspond to data recorded from, e.g.~different participants in
an experiment, or different time periods from the same participant. The whole
dataset from the first condition is denoted as $\bm x^N$, where $N$ is the
population size of these data, and the $N$ time series segments within $\bm
x^N$ as $\bm x_1$, $\bm x_2,\dots,\bm x_N$. Similarly, for the second condition
one has $\bm y^M = \{\bm y_1,\dots, \bm y_M\}$. Our goal is to decompose the
difference in $f$ between $\mathcal{X}$ and $\mathcal{Y}$ into spectral, phasic
and spectrum-phase interaction components -- i.e.~to decompose
\begin{equation}
\Delta\big(\bm x^N, \bm y^M\big) : = \bar{f}\big(\bm x^N\big) - \bar{f}\big(\bm y^M\big)~,
\end{equation}
where $\bar{f}\big(\bm x^N\big) = \frac{1}{N}\sum_{j=1}^N f\big(\bm x_j\big)$
and $\bar{f}\big(\bm y^M\big) = \frac{1}{M}\sum_{k=1}^M f\big(\bm y_k\big)$ are
the empirical ensemble averages of the function in question, $f$. This is
achieved by a series of comparisons between expected $f$ values on the data and
those on a set of progressively more constrained null models for the stochastic
processes underlying the data.

Formally, we consider $\bm x_1,\dots,\bm x_N$ to be independent and identically
distributed (i.i.d.) realisations of a stochastic process sampled under
condition $\cX$, and $\bm y_1,\dots,\bm y_M$ to be i.i.d. realisations of
another stochastic process sampled under condition $\cY$, and $\bm x_j,\bm
y_k\!\in\!\mathbb{R}^T$, where $T$ is the length of each time series. The
decomposition utilises the discrete Fourier transform, which is denoted by
$\hat{\bm x}=\mathcal{F}\{ \bm x\}\in\mathbb{C}^T$, given a time series $\bm
x$. The amplitudes of the Fourier components are denoted by $\bm A (\hat{\bm
x})=\{A_1(\hat{\bm x}),\ldots,A_T(\hat{\bm x})\} \in \mathbb{R}^T$, and their
phases by $\bm \phi (\hat{\bm x})=\{ \phi_1 (\hat{\bm x}),\ldots,\phi_T
(\hat{\bm x})\}\in [0,2\pi]^T$. Thus, the data for $\mathcal{X}$ can be
represented in the frequency domain as i.i.d. phase-amplitude tuples $\big( \bm
A(\hat{\bm x}_j) , \bm \phi(\hat{\bm x}_j) \big)$, following a distribution
$p_{\mathcal{X}}(\bm A, \bm \phi)$ induced by $\mathcal{X}$ -- and similarly
for the $\bm y_k$.

We begin by considering a null model $\mathcal{M}^\text{i}$ on which amplitudes
and phases have no interaction -- i.e. are statistically independent.
Accordingly, we construct new time series $\bm x_j^\text{(w)}$ that satisfy
this null model by combining the spectrum of each $\bm x_j$ with the phases
from some other randomly chosen time series from \emph{within} condition
$\mathcal{X}$ (and similarly for the $\bm y_k$). That is, we construct $\bm
x_j^\text{(w)} = \mathcal{F}^{-1}\{ \bm A(\hat{\bm x}_j) e^{i \bm \phi(
\hat{\bm x}_{\alpha_j})} \}$ and $\bm y_k^\text{(w)} = \mathcal{F}^{-1}\{ \bm
A(\hat{\bm y}_k) e^{i \bm \phi( \hat{\bm y}_{\beta_k})} \}$, where $\alpha_j$
and $\beta_k$ are distributed uniformly over $\{1,\dots,N\}$ and
$\{1,\dots,M\}$, respectively. We then consider the mean value of $f$ on these
phase-shuffled data, given by $\nu^{\text{i}}\big(\bm x^N\big) := \frac{1}{N}
\sum_{j=1}^N f\big(\bm x_j^\text{(w)}\big)$. The spectrum-phase interaction
contribution to the value of $f$ in condition $\mathcal{X}$ is then calculated
as
\begin{equation}
\Delta^\text{i}\big(\bm x^N\big) \coloneqq 
\bar{f}\big(\bm x^N\big) - \E\big\{ \nu^\text{i}\big(\bm x^N\big) \big| \bm x^N \big\}~,
\end{equation}
where the conditional expectation averages the effect of the random integers
$\alpha_j$ on $\nu^\text{i}$. Similarly, $\Delta^\text{i}\big(\bm y^M\big)$ can
be calculated for $\mathcal{Y}$. When estimating $\Delta^\text{i}(\bm x^N)$ and
$\Delta^\text{i}(\bm y^M)$ in practice, one will approximate the distribution
of $\nu^\text{i}$ by averaging multiple realisations of it.

The quantity $\Delta^\text{i}\big(\bm x^N\big)$ measures the extent to which
the expected value of $f$ would be affected if one were to break any dependence
that exists between the amplitudes and phases of the $\hat{\bm x}_j$.
Equivalently, $\Delta^\text{i}\big(\bm x^N\big)$ accounts for the deviation in
the mean value of $f$ in condition $\mathcal{X}$ from that which would be
expected if the null model $\mathcal{M}^\text{i}$ holds. For large $N$, the law
of large numbers guarantees that
\begin{align*}
  \Delta^\text{i}(\bm x^N) \!\to
  \E_{p_\mathcal{X}(\bm A,\bm \phi)}\{f(\bm A,\bm\phi)\} - \E_{p_\mathcal{X}(\bm
  A)p_\mathcal{X}(\bm\phi)}\{f(\bm A,\bm\phi)\},\nonumber
\end{align*}
and hence that in the absence of any dependency between the phases and spectra,
i.e.~when $p_\mathcal{X}(\bm A,\bm\phi) = p_\mathcal{X}(\bm
A)p_\mathcal{X}(\bm\phi)$, $\lim_{N\to\infty} \Delta^\text{i}\big(\bm
x^N\big)=0$.

Next, we focus on the \emph{phasic effect} on $f$, i.e.~the effect of
differences between the phase distributions of \cX and \cY. For this, we
consider a second null model $\mathcal{M}^\phi$ under which phases are not only
independent from amplitude but also follow the same distribution in each of the
conditions $\cX$ and $\cY$. We construct phase-shuffled time series $\bm
x_j^\text{(a)}, \bm y_k^\text{(a)}$ that satisfy this null model by replacing
the phases of each time series with those from another randomly chosen time
series from the whole set of data $\{\bm x^N, \bm y^M$\}. That is, we construct
$\bm x_j^\text{(a)} = \mathcal{F}^{-1}\{ \bm A(\hat{\bm x}_j) e^{i \bm \phi(
\hat{\bm w_j})} \}$ and $\bm y_k^\text{(a)} = \mathcal{F}^{-1}\{ \bm A(\hat{\bm
y}_k) e^{i \bm \phi( \hat{\bm z_k})} \}$, where $\hat{\bm w}_j,\hat{\bm z}_k$
are the discrete Fourier transforms of independently randomly chosen time
series that are each drawn from $\cX$ with probability 1/2, and from $\cY$ with
probability 1/2. Then, we consider the mean value of $f$ on these
phase-shuffled data: $\nu^{\phi}\big(\bm x^N|\bm y^M\big) := \frac{1}{N}
\sum_{j=1}^N f\big(\bm x_j^\text{(a)}\big)$, and introduce
\begin{align*}
\Delta^\phi\big(\bm x^N\big) &\coloneqq 
\E\Big\{ \nu^\text{i}\big(\bm x^N\big) - \nu^\phi\big(\bm x^N|\bm y^M\big) \Big| \bm x^N,\bm y^M \Big\}\,.
\end{align*}
We define $\Delta^\phi\big(\bm y^M\big)$ analogously. Again, when estimating
these quantities in practice, one can approximate the distributions of
$\nu^\text{i}$ and $\nu^\phi$, for each condition, by averaging multiple
realisations of them.

The quantity $\Delta^\phi\big(\bm x^N \big)$ measures the expected effect on
the mean value of $f$ in condition $\mathcal{X}$ if $\mathcal{M}^\phi$ holds --
i.e. the effect of changing the probability distribution of the phases from
$p_\mathcal{X}(\bm \phi)$ to the mixture $\left(p_\mathcal{X}(\bm \phi) +
p_\mathcal{Y}(\bm \phi)\right)/2$. Note that if the distribution of phases is
the same for both conditions, so that $p_\cX(\bm\phi)=p_\cY(\bm\phi)$, then the
law of large numbers guarantees that
\begin{equation}
\lim_{N\to\infty}\Delta^\phi\big(\bm x^N\big ) = \lim_{M\to\infty}\Delta^\phi\big(\bm y^M\big ) = 0~.    
\end{equation}

Finally, we consider the effect of spectral differences between the conditions
on the difference in $f$. For this, we consider the deviation of the
phase-shuffled data above from a further constrained null model
$\mathcal{M}^\text{A}$, in which both amplitudes and phases are statistically
independent and distributed identically in $\cX$ and $\cY$. Specifically, we
consider $\nu^{\text{A}}\big(\bm x^N,\bm y^M\big) := \nu^\phi\big(\bm x^N|\bm
y^M\big) - \nu^\phi\big(\bm y^M|\bm x^N\big)$. Since $\bm x_i^\text{(a)}$ and
$\bm y_i^\text{(a)}$ have, by definition, the same phase statistics,
$\nu^\phi\big(\bm x^N|\bm y^M\big)$ and $\nu^\phi\big(\bm y^M|\bm x^N\big)$
will, on average, differ only because of differences between the distribution
of the spectrum of $\mathcal{X}$ and $\mathcal{Y}$. Therefore, we introduce
\begin{align*}
\Delta^\text{A}\big(\bm x^N,\bm y^M\big) 
\coloneqq 
\E\big\{ \nu^{\text{A}}\big(\bm x^N, \bm y^M\big) \big| \bm x^N,\bm y^M \big\}
\end{align*}
as a metric of the spectral effect. If the distribution of the spectrum is the
same for both conditions, then $\lim_{N,M\to\infty}\Delta^\text{A}\big(\bm
x^N,\bm y^M\big) = 0$ by the law of large numbers. Again, when estimating this
quantity in practice, one will approximate the distribution of
$\nu^{\text{A}}\big(\bm x^N,\bm y^M\big)$ by obtaining multiple realisations.

\begin{figure*}[ht!]
  \centering
  \includegraphics{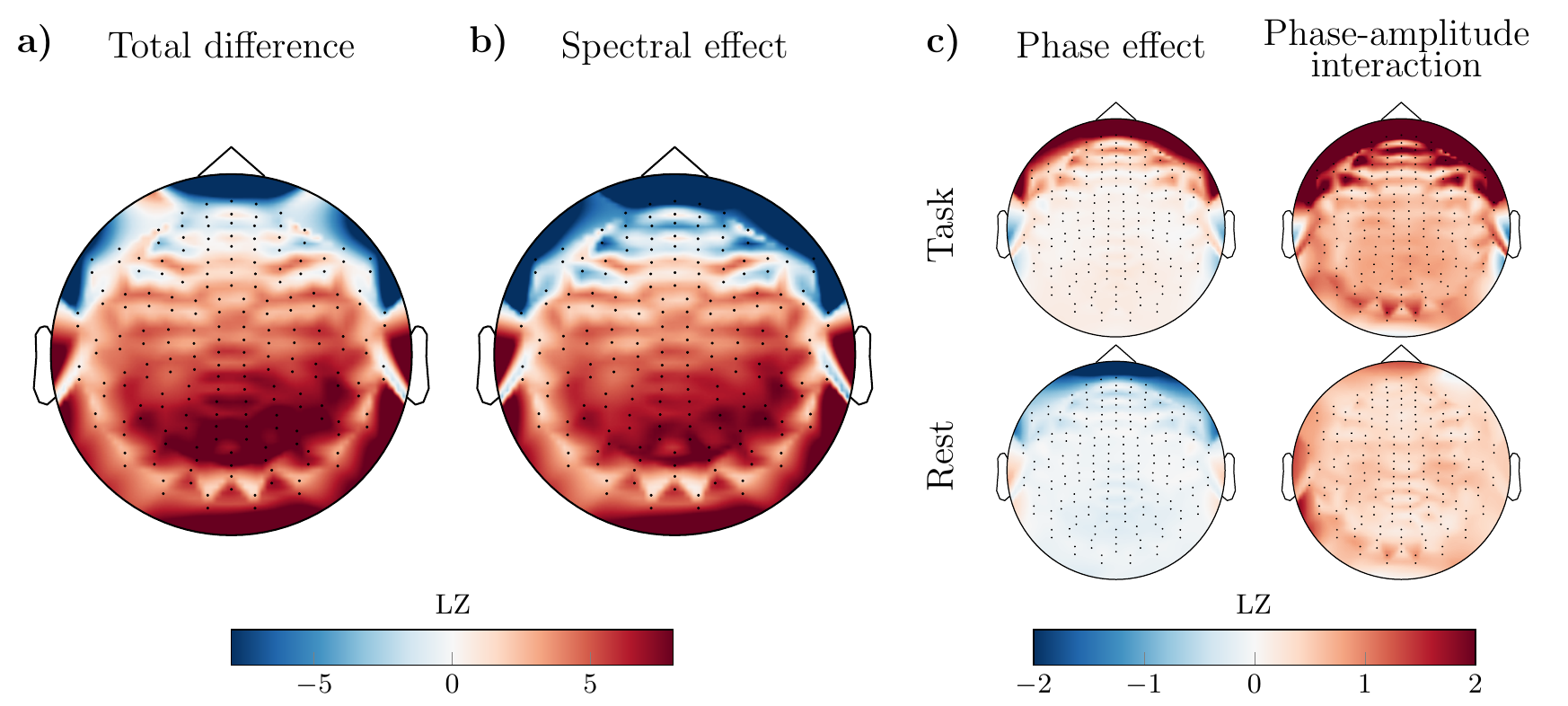}

  \caption{\textbf{Decomposition of the difference in Lempel-Ziv (LZ)
    complexity between task and rest conditions in the CAMCAN MEG dataset}.  The LZ complexity computed on sensor-level data during task minus that during rest indicates a pronounced
  reduction in complexity in frontal areas during task (\textbf{a}).
  This effect is mostly driven by spectral
  changes, $\Delta^\text{A}\big(\bm x^N, \bm y^M\big)$ (\textbf{b}). Nevertheless, the decomposition \eqref{eq:decomp} also reveals substantial
  differences in phase and phase-amplitude interaction contributions between conditions
  (\textbf{c}). These have a different spatial profile and show the
  opposite trend to the spectral component (top row shows $\Delta^\phi\big(\bm x^N\big)$ and $\Delta^\text{i}\big(\bm x^N\big)$ and bottom row shows $\Delta^\phi\big(\bm y^M\big)$ and $\Delta^\text{i}\big(\bm y^M\big)$). Due to the large number
  of participants, each quantity is significantly non-zero for most channels (i.e.~t-test across participants gives $p \ll 0.05$).
  }

  \label{fig:sng}
\end{figure*}

With these quantities at hand, via a telescopic sum we can obtain a
decomposition of the total difference in $f$ between the two conditions into
spectral, phasic, and spectrum-phase interaction terms. We have that the
difference in mean $f$ values between the conditions decomposes into
\def\hsep{3.35cm}
\begin{align}
\begin{array}{ll}
  \Delta(\bm x^N,\bm y^M)\!=\!%\makebox[\hsep][l]{=}
  & \begin{rcases}
    \makebox[\hsep][l]{$\Delta^\text{A}(\bm x^N, \bm y^M)$}
  \end{rcases} \text{Spectrum} \\%\vspace{5pt} \\
  & \multicolumn{1}{l}{
    \begin{rcases}
      \makebox[\hsep][l]{+$\Delta^\phi(\bm x^N) - \Delta^\phi(\bm y^M)$} \\
    \end{rcases} \text{Phase} %\vspace{5pt}
  } \\
  & \multicolumn{1}{l}{
    \begin{rcases}
      \makebox[\hsep][l]{$+\Delta^\text{i}(\bm x^N) - \Delta^\text{i}(\bm y^M)$\,} \\
    \end{rcases} \text{Interaction}
  }
\end{array}
\label{eq:decomp}
\end{align}

Of these, the first term is the difference in $f$ that persists on data
modified so the phases have the same distribution across conditions, and so
corresponds to the difference attributable to spectral changes only. Similarly,
by comparing the data with the observed phase distributions against data with
identically distributed phases, the second term measures the difference in $f$
attributable to phase changes. Finally, the third term compares the observed
data with phase-shuffled time series to account for changes due to the
phase-spectrum interaction in both conditions.

Accordingly, each of the $\Delta$'s can be considered to be comparing expected
$f$ values on the data against $f$ values on a set of increasingly restrictive
null models, $\mathcal{M}^\text{i} \to \mathcal{M}^\phi \to
\mathcal{M}^\text{A}$. We note that this decomposition is invariant to the
order in which the decomposition is constructed, i.e.~it doesn't make a
difference if phasic effects are considered before spectral contributions (as
described here), or vice versa (proof in Supp. Mat.).

\paragraph{Example.}

\footnotetext[999]{In these references, all channels are concatenated before
computing $C(\bm x)$. The version used here is what is commonly referred to as
``LZs'' in the neuroscience literature.}

As an illustration, we present an analysis of the entropy rate of binarised
magnetoencephalographic (MEG) signals, as measured with LZ complexity. We use
the Cambridge Centre for Ageing and Neuroscience (CAMCAN)
dataset~\cite{Shafto2014}, which includes a large-scale MEG dataset of
participants undergoing several cognitive tasks, and study the differences in
Lempel-Ziv complexity~\cite{Lempel1976} between participants in wakeful rest,
and participants performing a simple cognitive stop/no-go
task~\cite{Shafto2014}. This measure (or minor variations of it) has been
widely used in the neuroscience
literature~\cite{Zhang2001,Schartner2017a,Dolan2018,Note999}, showing a
remarkable performance in discriminating between different states of
consciousness, for instance normal wakefulness versus sleep~\cite{Casali2013}.

In this application, we consider data for 131 participants in both ``task'' and
``rest'' conditions. The data from each participant were divided up into 100
non-overlapping windows of length $T=1024$ (which corresponds to approximately
\SI{4}{\second} given the sampling rate of \SI{250}{\hertz}). To compute the LZ
complexity, time series were binarised, and then the original (1976) version of
the LZ complexity described in Ref.~\cite{Lempel1976} was computed.
Binarisation was carried out based on the mean value of the time series in
question, so the binarised time series contained ones where the raw value was
greater than the mean, and zeros where the raw value was less than the mean.

For each of the 204 MEG channels of each participant, the decomposition in
Eq.~\eqref{eq:decomp} was applied considering $\bm x^N$ to be the windowed data
during task and $\bm y^M$ to be the windowed data during rest, and using 500
realisations of the random variables involved (i.e. 500 random phase
shufflings). Thus, a set of $\Delta$'s was obtained for each channel, for each
participant. Then, to assess whether differences were significant at the group
level, 1-sample t-tests were carried out across participants --- for each of
the $\Delta$'s, for each channel. The mean value of each of the $\Delta$'s at
each MEG channel is shown in Fig.~\ref{fig:sng}.

Our decomposition reveals information about the relation between task and rest
that is not captured by other statistical tools. First, by studying the direct
difference between LZ complexity in task versus rest, our results show a
reduction of complexity in frontal regions, and an increase in the rest of the
brain during the task (Fig.~\subref{fig:sng}{a}). Our decomposition shows that
the vast majority of this difference (approximately 7.5 out of 8 units) can be
explained by spectral effects (Fig.~\subref{fig:sng}{b}). Interestingly,
contrasting effects are found in the phase and interaction components. In
particular, during task there is a strong and heavily localised phase-amplitude
interaction component, which becomes much weaker and spatially homogeneous
during rest (Fig.~\subref{fig:sng}{c}). Interestingly, both of these show the
\emph{opposite} trend from the direct difference, with an increase in frontal
regions and reductions elsewhere during task. The neurobiological implications
of these findings will be developed in a separate publication.

\paragraph{Conclusion.}

In this paper we have tackled the problem of determining to what extent a
measured difference in some quantity between two time series datasets can be
attributed to differences between their power spectra. For this, we introduced
a decomposition that uses a sequence of null models to disentangle the effect
of spectral, phasic, and phase-amplitude interaction effects. Our decomposition
requires no assumptions on the data (beyond that distinct samples within the
data are independent), and is easy to compute. As a proof of concept, we
provided an example of the decomposition yielding novel results on some
neuroimaging data, more nuanced than what was previously possible with a
standard analysis of LZ complexity.

Since this decomposition can be applied to any observed difference between two
datasets, it promises to be a valuable tool for practitioners in multiple
scientific disciplines. Moreover, it will help to deepen our understanding of
the behaviour of non-linear properties on datasets describing complex systems.
\\

The authors thank Lionel Barnett and Anil Seth for valuable discussions, and
two anonymous referees for comments on earlier versions of this manuscript. We
also thank Aleksi Ikkala and Darren Price for vital background work, and Yike
Guo for supporting this research. P.M. and D.B. are funded by the Wellcome
Trust (grant no. 210920/Z/18/Z). F.R. is supported by the Ad Astra Chandaria
foundation. D.B. conceptualised the work. A.B.B. guided the writing of the
paper.

\end{document}

\title[Decomposing spectral and phasic differences in non-linear features between datasets]{Supplementary Material to\\[2ex]Decomposing spectral and phasic differences in non-linear features between datasets}

\author{Pedro A.M. Mediano}
\thanks{P.M. and F.R. contributed equally to this work.\\E-mail: pam83@cam.ac.uk, f.rosas@imperial.ac.uk}
\affiliation{Department of Psychology, University of Cambridge, Cambridge CB2 3EB}

\author{Fernando E. Rosas} 
\thanks{P.M. and F.R. contributed equally to this work.\\E-mail: pam83@cam.ac.uk, f.rosas@imperial.ac.uk}
\affiliation{Center for Psychedelic Research, Department of Medicine, Imperial College London, London SW7 2DD}
\affiliation{Data Science Institute, Imperial College London, London SW7 2AZ}
\affiliation{Center for Complexity Science, Imperial College London, London SW7 2AZ}

\author{Adam B. Barrett}
\thanks{A.B.B. and D.B. are joint senior authors.}
\affiliation{Sackler Center for Consciousness Science, Department of Informatics, University of Sussex, Brighton BN1 9RH}

\author{Daniel Bor}
\thanks{A.B.B. and D.B. are joint senior authors.}
\affiliation{Department of Psychology, University of Cambridge, Cambridge CB2 3EB}

% \date{\today}

\maketitle

\section{Naive applications of conventional surrogate methods fail at two-sample comparisons}

Let us start by considering the problem of testing if an observable of interest
$f(\bm x) \in\mathbb{R}$ depends only on the power spectrum of a single dataset
composed by the time series $\bm x=(x_1,\dots,x_T) \in \mathbb{R}^T$. The
discrete Fourier transform of $\bm x$ is denoted as $\hat{\bm
x}=\mathcal{F}\{\bm x\}\in \mathbb{C}^T$, with amplitude $A(\hat{\bm x})= |
\hat{\bm x} | \in\mathbb{R}^T$ and phase $\phi(\hat{\bm x}) =
\arctan\big(\mathfrak{Im}( \hat{\bm x}) / \mathfrak{Re}(\hat{\bm x}) \big) \in
[0,2\pi]^T$. A simple procedure, known as \emph{phase
randomisation}~\cite{Theiler1992}, is to compare the value of $f(\bm x)$
against a null distribution given by the random variable $f(\bm
x_{\text{pr}})$, where $\bm x_{\text{pr}}$ is surrogate data obtained by taking
the Fourier transform of $\bm x$, adding an independent random phase to each
component, and then taking the inverse Fourier transform. Hence, $\bm
x_{\text{pr}} = \mathcal{F}^{-1} \{ A(\hat{{\bm x}}) e^{i {\bm \phi} }\} $
where $\bm \phi$ is a random vector of uniformly distributed phases, denoted by
$\bm\phi\sim \mathcal{R}$. (Technically, half of the entries of $\bm\phi$ are
uniformly distributed over $[0,2\pi]$ while the other half are their complex
conjugates, so that $\bm x_\text{pr}$ is a real
vector~\cite{bracewell1986fourier}.) Accordingly, the null hypothesis that
there is no genuine non-linear structure is rejected if the quantity $f(\bm x)
-\E\{f(\bm x_{\text{pr}})\}$ is significantly different from zero~\footnote{It
is well known that this test corresponds to comparing $\bm x$ against time
series generated by a linear model with the same autocorrelation as $\bm x$.}.

Let us now consider a slightly more complex scenario with two
time series $\bm x$ and $\bm y$, and consider whether the difference $ \delta
\coloneqq f(\bm x) - f(\bm y)$ can be attributed solely to differences in their
spectra. A naive approach to address this problem -- considered here for
illustration purposes -- would be to compare $\delta$ against the null
distribution that corresponds to the random variable
\begin{align}
  \tilde{\delta} &= f(\bm x_{\text{pr}}) - f(\bm y_{\text{pr}}) ~,
  \label{eq:indep_shuffle}
\end{align}
where $\bm x_\text{pr} = \mathcal{F}^{-1} \{ A(\hat{\bm x}) e^{i\bm\phi_1}\}$
and $\bm y_\text{pr} = \mathcal{F}^{-1} \{ A(\hat{\bm y}) e^{i\bm\phi_2}\}$,
with $\bm \phi_1,\bm\phi_2\sim\mathcal{R}$ being statistically independent of
$\hat{\bm x}$, $\hat{\bm y}$, and of each other. One would then reject the null
hypothesis if $\delta - \E\{\tilde\delta\}$ is significantly different from
zero. This is equivalent to testing the difference between the ``corrected''
values $f(\bm x) - \E\{ f( \bm x_\text{pr} ) \}$ and $f(\bm y) - \E\{ f( \bm
y_\text{pr} ) \}$.

\begin{figure}[ht!]
  \centering
  \includegraphics{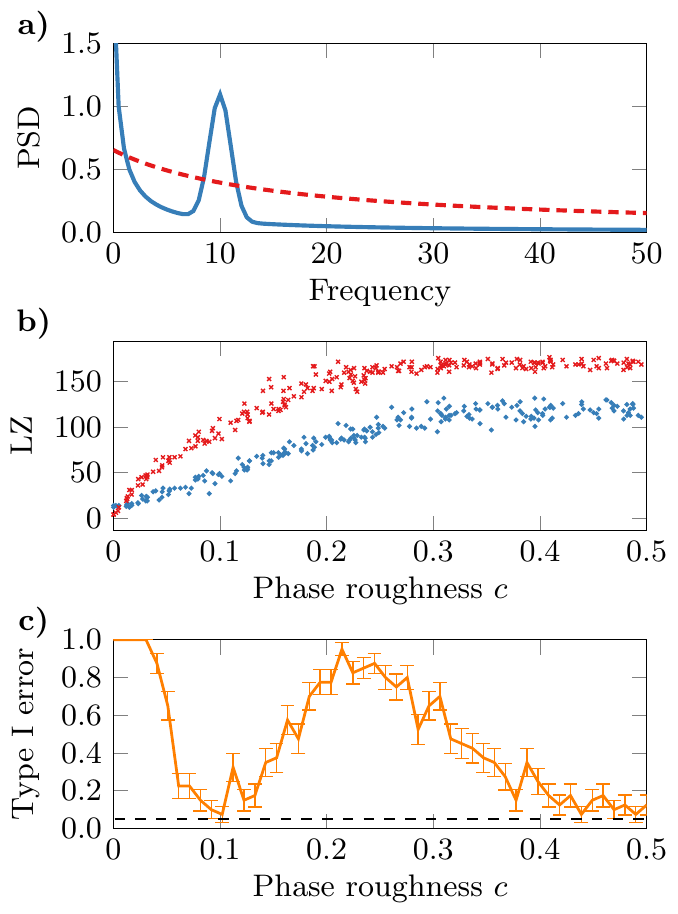}

  \caption{\textbf{Naive application of typical surrogate methods to two-sample
comparisons results in false positives}. \textbf{a)} Two spectra, shown in
\textcolor{BrewerBlue}{blue} (solid) and \textcolor{BrewerRed}{red} (dashed),
were used to generate time series via inverse Fourier transform. \textbf{b)} To
generate phases we used a simple model with a roughness parameter $c$, that
interpolates between constant phase ($c=0$) and fully random phases ($c=1$).
The Lempel-Ziv complexity~\cite{Lempel1976} of signals
generated with both spectra and the same phase is shown in
\textcolor{BrewerBlue}{blue} (bottom) and \textcolor{BrewerRed}{red} (top).
\textbf{c)} Naive applications of typical
surrogate methods, like the one described in 
Eq.~\eqref{eq:indep_shuffle}, incorrectly
reject the null hypothesis when applied to two time series with identical
phases but different spectra.}

  \label{fig:roughness}
\end{figure}

Unfortunately, this test generally fails because the difference
between two null models is not necessarily a good null model of the difference. 
More specifically, the differences $\tilde{\delta}$ 
that are seen when injecting
randomised phases may not be representative of
the differences that are seen when injecting phases from other (plausibly more
realistic) distributions.

To illustrate these ideas, let's consider two given power spectra, shown in
Fig.~\subref{fig:roughness}{a}, and generate random phases via a simple model
equipped with a roughness parameter $c\in[0,1]$ that interpolates between a
constant phase ($c=0$) and fully random phases ($c=1$)~\footnote{The details of
the phase-generating model are irrelevant -- it only matters that there is a
method for generating phases that are unlike samples from \R when $c<1$.}. With
this, we can build $(\bm x_c,\bm y_c)$ pairs by combining the two spectra in
Figure~\subref{fig:roughness}{a} and \emph{the same phase} $\bm\phi_c$
generated with a given $c\in[0,1]$. The null hypothesis is true (by
construction) for all these pairs, as $\bm x_c$ and $\bm y_c$ only differ in
their spectra. However, as shown in Fig.~\subref{fig:roughness}{c}, the test in
Eq.~\eqref{eq:indep_shuffle} results in a large number of false positives.

To understand this result, note that the value of $\E\{ f(\bm x_c) - f(\bm
y_c)\}$ (i.e. the expected value of $\delta$ under the null hypothesis) roughly
corresponds to the gap between the point clouds in
Fig.~\subref{fig:roughness}{b}. However, the test in
Eq.~\eqref{eq:indep_shuffle} is related to the above quantity when $c=1$. The
crux is that the two spectra ``max out'' at different values of $f$, making the
quantity $\E\big\{ \tilde{\delta}
\big\}$ non-zero and, in turn, introducing a bias in the test for $c<1$.

\section{The decomposition is invariant to ordering of the null models}

Here we prove that our proposed decomposition does not depend on the ordering
in which phasic and spectral effects are considered. In particular, instead of
the studied sequence of null models given by
$\mathcal{M}^\text{i}\to\mathcal{M}^\phi\to\mathcal{M}^\text{A}$, one could
also consider
$\mathcal{M}^\text{i}\to\mathcal{M}^{\text{A}'}\to\mathcal{M}^{\phi'}$ where
the effect of spectrum is considered before the effect of phase. In the rest of
this section, we prove that this second sequence of null models gives the same
decomposition.

For this purpose, let us define $\bm x_j^\phi = \mathcal{F}^{-1}\{ \bm
A(\hat{\bm u}_j) e^{i \bm \phi( \hat{\bm x_j})} \}$ and $\bm y_k^\phi =
\mathcal{F}^{-1}\{ \bm A(\hat{\bm v}_k) e^{i \bm \phi( \hat{\bm y_k})} \}$,
where $\hat{\bm u}_j,\hat{\bm v}_k$ are the discrete Fourier transforms of
independently randomly chosen time series that are each drawn from $\cX$ with
probability 1/2, and from $\cY$ with probability 1/2. With this, let us
introduce $\nu^{\text{A}'}(\bm x^N|\bm y^M) := \frac{1}{N} \sum_{j=1}^N f(\bm
x_j^\phi)$, and define
\begin{align*}
\Delta^{\text{A}'}(\bm x^N) &\coloneqq 
\E\Big\{ \nu^\text{i}(\bm x^N) - \nu^\text{A}(\bm x^N|\bm y^M) \Big| \bm x^N,\bm y^M \Big\}
\end{align*}
and $\Delta^{\text{A}'}(\bm y^M)$ analogously. Hence, the difference
$\Delta^{\text{A}'}(\bm x^N, \bm y^M) \coloneqq \Delta^{\text{A}'}(\bm x^N) -
\Delta^{\text{A}'}(\bm y^N)$ is the spectral component of the difference when
assessed on the second stage of the alternative decomposition given by
$\mathcal{M}^\text{i}\to\mathcal{M}^{\text{A}'}\to\mathcal{M}^{\phi'}$.

\begin{proposition}
The decomposition given by \begin{equation} \mathcal{D}_1:
\mathcal{M}^\textnormal{i}\to\mathcal{M}^\phi\to\mathcal{M}^\textnormal{A}
\end{equation}
is equivalent to the decomposition given by
\begin{equation}
\mathcal{D}_2: \mathcal{M}^\textnormal{i}\to\mathcal{M}^{\textnormal{A}'}\to\mathcal{M}^{\phi'}~.
\end{equation}
\end{proposition}
\begin{proof}

Because both decompositions start with $\mathcal{M}^\textnormal{i}$, it is
clear that the term correposponding to spectral-phasic interaction is
equivalent. Therefore, it is enough to show that either the spectral or the
phasic contribution is the same, as proving one would imply the other. Our
strategy is to prove that the spectral component assessed in the third step in
$\mathcal{D}_1$, given by $\Delta^\text{A}(\bm x^N, \bm y^M)$, is equal to the
spectral component estimated in the second stage of $\mathcal{D}_2$ given by
$\Delta^{\text{A}'}(\bm x^N, \bm y^M)$.

To prove this, let us introduce $\bm x_j^\text{y} = \mathcal{F}^{-1}\{ \bm
A(\hat{\bm x}_j) e^{i \bm \phi( \hat{\bm y}_\beta)} \}$ and $\bm y_k^\text{x} =
\mathcal{F}^{-1}\{ \bm A(\hat{\bm y}_k) e^{i \bm \phi( \hat{\bm x}_\alpha)}
\}$, where $\alpha, \beta$ are integers sampled at random from $\{1,\dots,N\}$
and $\{1,\dots,M\}$, respectively. Put simply, $\bm x_j^\text{y}$ and $\bm
y_k^\text{x}$ have the spectrum of one process combined with a randomly sampled
phase from the other. Furthermore, let us use the notation $\nu^\text{c}(\bm
x^N) = \frac{1}{N}\sum_{j=1}^N f(\bm x_j^\text{y}) $ and $\nu^\text{c}(\bm y^M)
= \frac{1}{M}\sum_{k=1}^M f(\bm y_k^\text{x})$. With this and the definition of
$\nu^\text{A}$, one can show that
\begin{align*}
\E\left\{ \nu^\text{A}(\bm x^N|\bm y^M) \Big| \bm x^N,\bm y^M \right\} = \E\left\{ \frac{\nu^\text{i}(\bm x^N) + \nu^\text{c}(\bm y^M)}{2} \Big| \bm x^N,\bm y^M \right\}~,
\end{align*}
and therefore
\begin{align*}
\Delta^{\text{A}'}(\bm x^N) = \E\left\{ \frac{\nu^\text{i}(\bm x^N) - \nu^\text{c}(\bm y^M)}{2} \Big| \bm x^N, \bm y^M \right\}~.
\end{align*}
An analogous calculation gives that
\begin{align*}
\Delta^{\text{A}'}(\bm y^M) = \E\left\{ \frac{\nu^\text{i}(\bm y^N) - \nu^\text{c}(\bm x^N)}{2} \Big| \bm x^N, \bm y^M \right\}~.
\end{align*}
Finally, combining all this one finds that
\begin{align*}
\Delta^{\text{A}'}(\bm x^N, \bm y^M)
=& \Delta^{\text{A}'}(\bm x^N) - \Delta^{\text{A}'}(\bm y^M)  \\
=& \E\Big\{ \frac{ \nu^\text{i}(\bm x^N) + \nu^\text{c}(\bm x^N)}{2} \Big| \bm x^N, \bm y^M \Big\} \\
& - \E\Big\{ \frac{\nu^\text{i}(\bm y^M) + \nu^\text{c}(\bm y^M)}{2} \Big| \bm x^N, \bm y^M \Big\} \\
=& \E\Big\{ \nu^\phi(\bm x^N|\bm y^M) - \nu^\phi(\bm
y^M|\bm x^N) \Big| \bm x^N, \bm y^M \Big\} \\
=& \Delta^\text{A}(\bm x^N, \bm y^M)~,
\end{align*}
which concludes our proof.
\end{proof}